\def\year{2019}\relax
\documentclass[letterpaper]{article} % DO NOT CHANGE THIS
\usepackage{aaai19}  % DO NOT CHANGE THIS
\usepackage{times}  % DO NOT CHANGE THIS
\usepackage{helvet} % DO NOT CHANGE THIS
\usepackage{courier}  % DO NOT CHANGE THIS
\usepackage[hyphens]{url}  % DO NOT CHANGE THIS
\usepackage{graphicx} % DO NOT CHANGE THIS
\def\UrlFont{\rm}  % DO NOT CHANGE THIS
\usepackage{graphicx}  % DO NOT CHANGE THIS
\title{Mobile community sensing with smallholder farmers in a developing nation;\\
A scaled pilot for crop health monitoring.}
\author{\Large \textbf{Daniel Mutembesa\textsuperscript{\rm 1}, Ernest Mwebaze\textsuperscript{\rm 1}, Solomon Nsumba\textsuperscript{\rm 1}}\\
\Large \textbf{Christopher Omongo}\textsuperscript{\rm 2}  \Large \textbf{Humphrey Mutaasa}\textsuperscript{\rm 3}\\ % All authors must be in the same font size and format. Use \Large and \textbf to achieve this result when breaking a line
\textsuperscript{\rm 1}Artificial Intelligence Research Lab, Makerere University.\\ 
mutembesa.daniel@gmail.com, emwebaze@google.com, snsumba@gmail.com % email address must be in 
\\
\textsuperscript{\rm 2}National Agricultural Research Organisation.\\ chrisomongo@gmail.com
\\
\textsuperscript{\rm 3}Uganda National Farmers Federation\\ mutaasah2@gmail.com}

\begin{document}

\maketitle

\begin{abstract}
Previously, crowdsourcing experiments in surveillance of crop diseases and pest have been trialed as small scale community sensing campaigns with select cohort of smallholder farmers, extension and experts.
While those pilots have demonstrated the viability of community sensing with mobile phones to collect massive amounts of real-time data all year round, to compliment low-resourced agricultural expert surveys, they are limited in generalising ideas for scaled implementations of a community sensing system with farmer communities.
\\
This work presents a case of scaled deployment of the mobile ad hoc surveillance for crowdsourcing real-time surveillance data on cassava from over 175 smallholder farmers across Uganda.
% based on the pilot run by MCROPS. 
This paper describes a modified mobile ad hoc surveillance ecosystem to suite smallholder farmer agents, a communication model and data collection model designed to cover the spatial interests for the scale of surveillance, a deployment plan, the training methodology and incentives structure. 
The paper also presents very early results of contributions from farmer agents, that could be usable in monitoring the movement of planting materials between districts, mapping cassava varieties, multiplication sites, and communities with little or no access to agricultural extension services, and possibly guide precision expert surveys in areas of high disease incidence. 
\end{abstract}

\section{Introduction}
%Crowdsourcing enables researchers to easily and quickly collect a large number of subjective ratings from a diverse set of participants. 
Crowdsourcing enables diverse set of participants easily contribute to subjects of interest to the requester.
Recently, a subset of crowdsourcing known as community sensing has gained ground in low-resource settings of the developing world due to the pervasive presence of mobile phone technology within remote rural communities.
This diversity of rural communities further enables research in cultural effects, influence factors generated by different %end-user devices, 
incentives, and effects of different surrounding environments. 
%However, these opportunities come at costs. Experimenters have only limited control of the test settings and in which environmental conditions the study takes place. Additionally, the remote test procedure can be error-prone as the test participants are not under the direct supervision of the test conductor. 
Extending a real-time mobile reporting toolkit to the agricultural actors across the country, while at the same time enabling experts to conduct surveillance tasks more effectively, is important for guiding timely reporting, precision in expert surveys and timely intervention planning to control viral cassava disease, vector and pest outbreaks.

\section{Related Work}
The work here majorly draws on insights and recommendations of that small scale pilot in (Mutembesa et. al., 2018). The paper presents a mobile ad hoc surveillance ran as a 20 month trial, using mixed select cohort of 29 volunteers including; smallholder farmers, experts and extension workers.
\\
The \emph{MCrops} project in that paper, ran a two year pilot (2016 to 2017) of a mobile ad hoc surveillance system that prototyped a small set up where a cohort of 29 farmer, extension and expert agents across the country contributed weekly reports to a cassava situation map. Farmers and extension agents would capture and upload geo-tagged images of diseased \emph{Cassava} leaf images, pests etc, in their localities to a central server then to a live interactive map.
The pilot's main objective was to find an optimal crowdsourcing scenario with farmers in disparate remote locations, the incentive considerations necessary for sustaining such efforts, and suggest recommendations for deployments of scale.
%\\\\
%A key component of this work is the ability to leverage the ubiquity of smartphone technology particularly in the African context. Mobile phone sensing offers a variety of novel, efficient ways to opportunistically collect data, enabling numerous mobile crowdsourced sensing (MCS) applications. Examples from alternate fields include (Mohan, Padmanabhan, and Ramjee 2008) and (Thiagarajan et al. 2009) where systems are developed for crowdsourcing traffic information. More related examples to our cassava disease sensing application include the use of crowds and communities in noise pollution sensing (Rana et al. 2010) and in air pollution sensing (Stevens and DHondt 2010). An even closer example of a system using different incentives to link credible buyers to sellers of market produce in Uganda is the Kudu system (Ssekibuule, Quinn, and Leyton-Brown 2013) which uses a double auction design algorithm to link buyers and sellers of agricultural produce.

\section{ \emph{AdSurv} mobile app and platform}
 Significant improvements had to be made to the already existing \emph{AdSurv} app presented in the work (Mutemebsa et al., 2018). The original app was a single function application with only one module for data collection. With the continued request for feedback mechanisms that are farmer-facing, three modules were added to the new mobile-based app as shown in Figures 1 and 2.
 
\subsection{Modified \emph{AdSurv} app}
The modifies \emph{AdSurv} app consists of four modules as follows;

\subsubsection{Data collection module}
The data collection module is described in detail in Mutembesa et al., 2018 and allows an agent to take a image of interest, geo-tag it with GPS coordinates and complete micro-tasks like labelling the image and writing observational comments. The agent can then upload the report to a server which maps it in real time to an online map.

\subsubsection{Diagnosis module}
The diagnosis module run a trained machine learning classification model that runs an optimised mobile neural network to identify the four most important diseases of Cassava namely; Cassava Mosaic Disease, Cassava Brown Streak Disease, Cassava Green Mite and Cassava Bacterial Blight. 
Cassava Brown Streak Disease and Cassava Green Mite are the most difficult to identify for smallholder farmers and most often require an expert to point them out.\\
The farmer opens the module, positions the camera over the leaf of interest, taps the screen once and is given a diagnosis within a few seconds.\\
A detailed presentation of the classifier is in [CITATION]. %%%%%%
%mobile neural network 

\subsubsection{News channel feed}
A news channel was added to the new \emph{AdSurv} app, which was supposed to provide regular online agriculture related content that is also accessible by the farmers offline.
The content desired by the farmers included articles on pest control, symptoms of diseases of interest,
announcements on new crop high yielding varieties, harvest handling, best agronomy practices, value addition to agricultural produce etc. 

\subsubsection{Chat and Question and Answer \emph{Q\&A}}
A unique module included in this community sensing scenario was a direct messaging chat application added to the \emph{AdSurv} app.
It allowed both smallholder farmers and experts registered to the app to directly communicate amongst themselves on matters related to cassava and agriculture, and exchange knowledge between one another.
The platform also quickly became a platform to gather insights on trends of interest for farmers as the seasons of cropping changed throughout the year of deployment.

\subsection{Equipment} 
Drawing from the lessons of previous work mentioned, the technical team noted that connectivity in rural areas is a challenge and one of the important factors to the success of mobile-based community sensing campaign.
 The two most important features of the phone to be considered were good connectivity capability relative to a battery life that lasted a whole day while within the USD 80 budget per phone.
%The desired features of the phones where good battery and good connectivity for areas that don’t have good connectivity in the rural area. 
Mobile smart phone with the following specifications were considered;
\begin{itemize}
   \item A full day's worth of battery; which minimises the worry for the farmer having to charge in the middle fo the day while in the farm.
   \item Processing power to run disease detection model; to run the diagnosis model without hanging and compromising accuracy of the model vis-a-vis the battery usage .
    \item Strong and reliable GPS was a mandatory for the chosen equipment. \\
    Previously, failure of GPS to resolve for some agents was a major demotivation factor. This was suspected to be a limitation with the phone equipment.
    %We received XXX number of Images without GPS in the pilot and was a complete failure for some agents causing major demotivation. 
    The equipment considered needed to be reliable in resolving GPS coordinates even in the remotest areas.
    \item Good connectivity in rural areas and reliable internet capabilities was key. Previously network issues were prevalent and a major concern in the first pilot, where agents collecting data out in the fields in some remote rural areas found it hard to upload their reports.
\end{itemize}

\begin{figure}[htbp]
\centerline{\includegraphics[width=8.5cm]{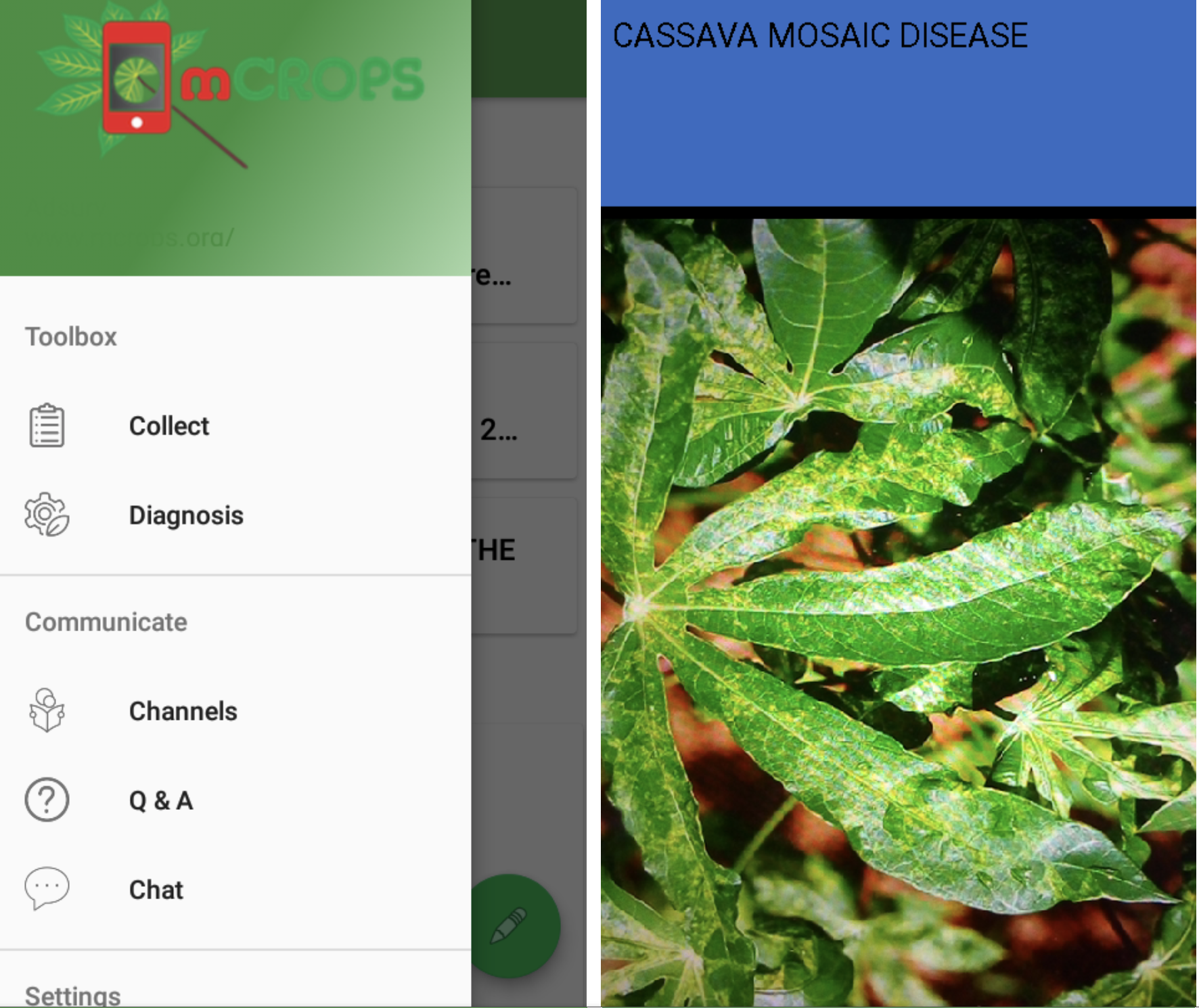}}
\caption{AdSurv app menu and the diagnosis module in action.}
\label{fig}
\end{figure}

\begin{figure}[htbp]
\centerline{\includegraphics[width=8.5cm]{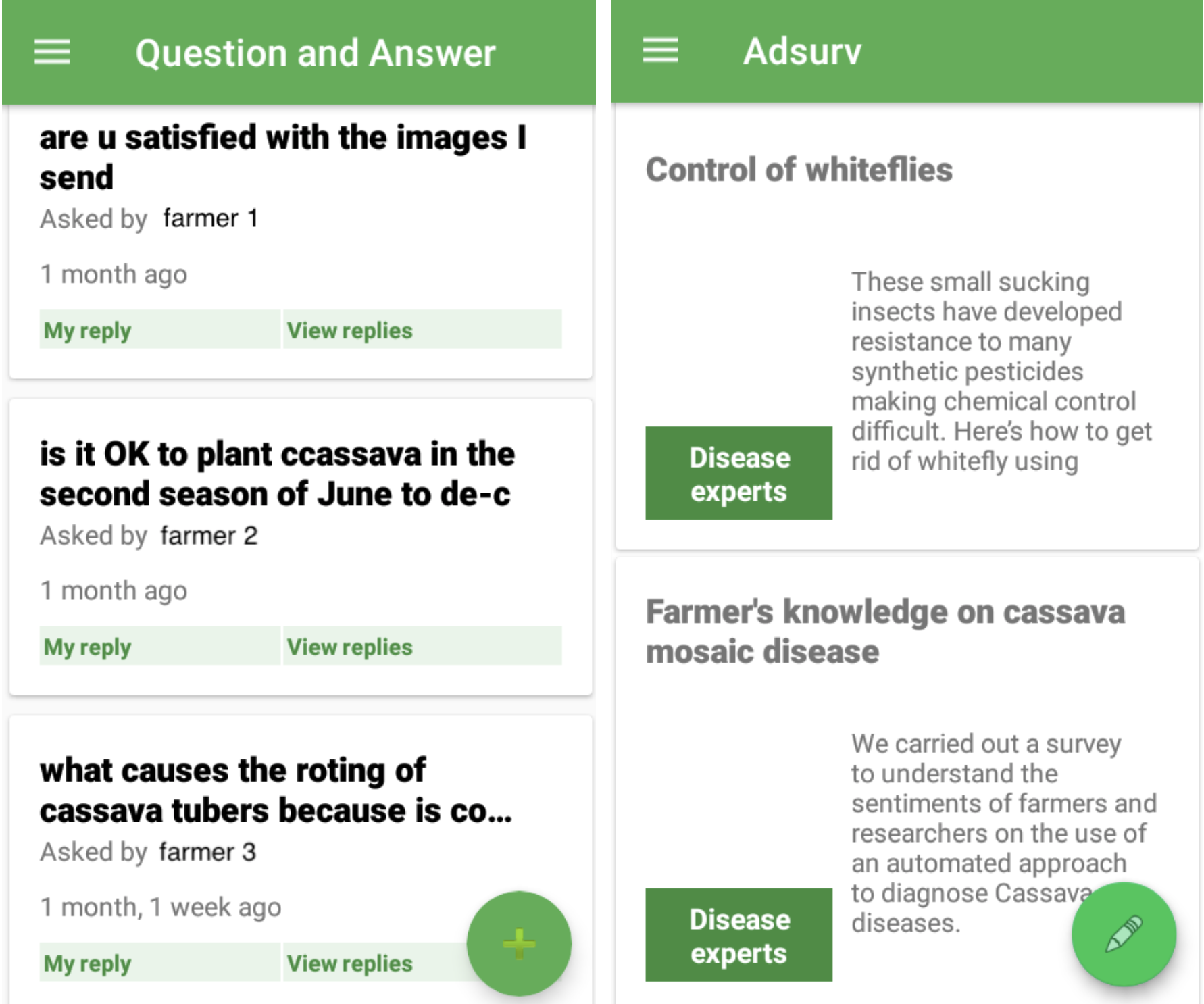}}
\caption{AdSurv app live chat and news feed modules.}
\label{fig}
\end{figure}

\section{A crowd selection model}
The cassava crop is of special interest as the most important food security crop in most households of Uganda and is a security food crop for the bigger part of sub Sahara n Africa. Its importance is captured in [citation]

\subsection {Agronomical zones of interest} 
Five agricultural zones of Uganda were chosen based on active cassava growing populous. Experts from the National Crops Resources and Research Institute (\emph{NaCRRI}) together with the Uganda National Farmers Federation (\emph{UNFFE}), contacted and verified 246 smallholder farmers from the five agricultural regions of Uganda. Each of the five regions was accorded a 40 phones for distribution out of the 200 phones commissioned for deployment, as per the selection criteria.
The five regions of interest are shown in Table 1.  %%%% figure 1
For the North of Uganda, six administrative districts were selected, five districts were selected for the West Nile region and ten districts for the greater North East region because it has the biggest population engaged in cassava production. Seven districts represented the eastern region, with six districts the central region of Uganda, and the ten districts selected for the western region of Uganda, split into two training cohorts; near West and a far West cohorts.\\
The South Western highlands were not included in this pilot because they support little of cassava growing activities.\\
Each region is represented by a Zonal Agricultural Research District Institute \emph{ZARDI}.

\subsection{Selection criterion for recruiting volunteer farmer crowds.} 
A selection criterion was developed and vetted along with the agriculture experts from \emph{NaCRRI} and \emph{UNFFE}. It drawing from lessons from the work of Mutembesa et.al 2018.
\begin{itemize}
   \item Each region was assigned 40 phones for distribution within the most active districts of each region.
   \item A selected farmer must be strictly registered as part of famers' group at district-level and an active member by the time of verification.
   \item A selected farmer should be in the age range : 20 years to 60 years.
    \item A selected should have attained at least primary school education level.
     \item A selected farmer must have a functional mobile phone, which means they already have know-how of use and mechanisms for maintenance for example charging battery, loading units for calls, using the phone for communication and performing purchases from mobile money platforms.
     \item A selected farmer must have an active mobile money account so as to be able to receive micro-reward payments over the phone.
     \item A selected agent must be a smallholder farmer i.e. 0.5 acres — 2 acres of cultivated land, not limited to mono crop but also multi crop known as mixed cropping. 
     \item Should be able to read and write.
     \item Must have access to mobile charging unit within community.
     \item Should be able to inform spouse or relative about the possession of phone to avoid wrangles.
     \item Should have a national ID or a letter from the local council chairperson. 
     \item A selection for each district should include try as much as possible to come up with two female farmers, two males and two youth members in the age range of 20 years to 30 years.\\\\
      %Gender parity \& representation for selection i.e half number of for each
    Exceptions were considered as follows;
    \item For regions with less number of verified farmers compared to others, reassignment was made to the next group or district or region with greater potential.
    \item For areas where either gender requirements would not meet the initial projections, redistribution would be to the next available farmers regardless of gender.
     \item For districts where we could not find smallholder farmers fitting the criteria requirements (two-six), we engaged small and medium scale cassava producers, who have greater acreage under cultivation.
\end{itemize}

\subsubsection{Sourcing for farmers.}
Finding and registering farmers were done through the networks of extension services workers of \emph{UNFFE} and \emph{NaCRRI}. They have presence and administrative representatives at the district level also known as a district farmer federation representative for \emph{UNFEE} and district agricultural production officer affiliated to \emph{NaCRRI}.\\
A composite list of 246 active cassava farmers was sourced for the verification exercise.
%The figure 2 shows the first selected farmers in COLOUR X

\subsection{Verification of small-holder farmers.}
A verification exercise was conducted along with the experts from \emph{NaCRRI and UNFFE}. 
Two teams were dispatched to follow the two routes; one west the North west also known as West Nile region and another to East through North and North Eastern region, to verify which of the 246 farmers sourced from the district representatives fitted the criteria of selection.\\
Justification for the verification process lay in the experiences of the first pilot (Mutembesa et. al., 2018), where three out of twenty-nine agents who were unverified, did not make any submission during the lifetime of the pilot.\\
The verification exercise following the selection criterion filtered 175 farmers out of the 246 identified.
The farmers that were not selected but fitted the selection criteria were put on the wait list incase a confirmed farmer was unable to join the project.
\\\\ %Form the lesson leaned in Mutembesa et. al 2018, try to Ensure the quality of participating crowd. 
Unlike other crowdsourcing scenarios where agents arrive online, in community sensing campaign as in this case image-based crop surveillance using mobile phones, where a crowd is pre-selected based on a referral basis of local extension service, it is vital to have mechanisms for improving the quality of crowd. 
\begin{itemize}
   \item These mechanisms are important for checking collusion within extension workers who source the farmers, to include themselves and unmeriting persons of their interest. 
   \item The mechanisms are vital for minimising the verification budget and the churn of unfitting farmers at the selection of crowd step.
   \item They help minimised possibility of attrition of both equipment and volunteers during the project.
\end{itemize}

The methodology is explained in detail and justified by [citation] that demonstrated working with trusted entities for rural-based crowds in participatory crop sensing campaigns

\begin{table}[htbp]
\caption{Farmer selection statistics}
\begin{center}
\begin{tabular}{|c|c|c|c|c|c|}
\hline
\textbf{Regional}&\multicolumn{5}{|c|}{\textbf{Selection statistics}} \\
\cline{2-6} 
\textbf{ZARDI} & \textbf{\textit{Region}}& \textbf{\textit{Districts}}& \textbf{\textit{Select}}& \textbf{\textit{M}}& \textbf{\textit{F}}\\
\hline
Arua & West Nile & 8 & 23 & 18 & 5 \\
\hline
Bulindi & Near West & 5 & 16 & 8 & 8 \\
\hline
Lira & North & 11 & 43 & 29 & 14 \\
\hline
Rwebitaba & Far West & 4 & 8 & 5 & 3 \\
\hline
Soroti & Far East & 10 & 32 & 20 & 12 \\
\hline
Tororo & Near East & 3 & 17 & 9 & 8 \\
\hline
Wakiso & Central & 7 & 36 & 23 & 13 \\
\hline
Totals &  & 48 & 175 & 112 & 63 \\
\hline
%\multicolumn{4}{l}{$^{\mathrm{a}}$Sample of a Table footnote.}
\end{tabular}
\label{tab1}
\end{center}
\end{table}

%\begin{figure}[htbp]
%\centerline{\includegraphics[width=8.5cm]{images/gender-selection.png}}
%\caption{Map of selected farmer per zonal regional agricultural research stations.}
%\label{fig}
%\end{figure}

\begin{figure}[htbp]
\centerline{\includegraphics[width=9.5cm]{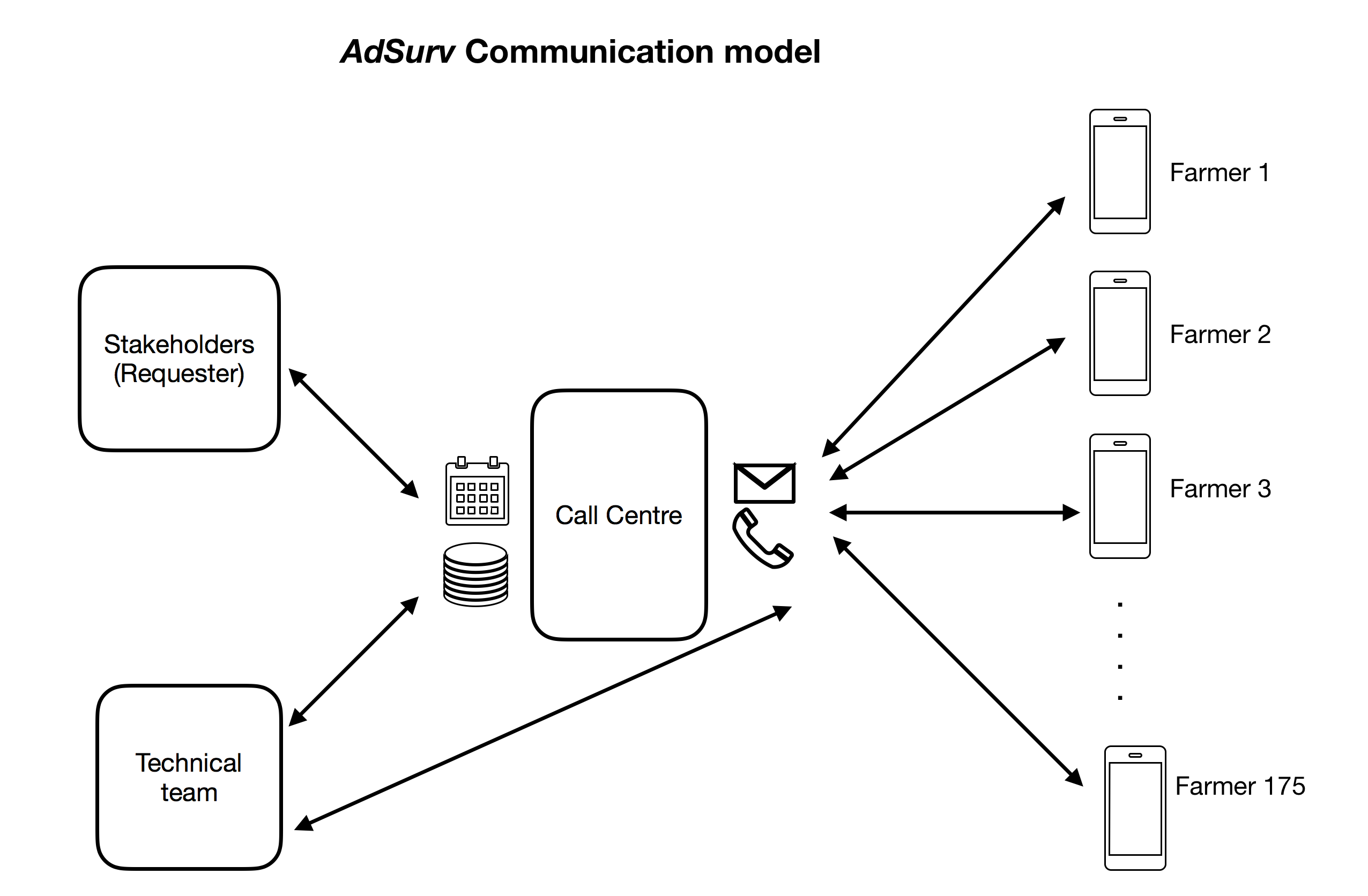}}
\caption{Call centre communication model}
\label{fig}
\end{figure}

\subsection{Call centre communication model}
Drawing from lessons in pilot done in (Mutembesa et al. 2018), where communication-type incentives yielded some desired participation, this scaled pilot followed recommendations and set up a call centre whose mandate was central communication channel to and from the farmers. 
A team of six was responsible for handling on average 29 farmers each.
\\
A centralised call centre handles the communication needs of the pilot. It broadcasts the messages and tasks from the requesters (stakeholder agencies) to the farmer network by both sms and regular calls. It also makes scheduled calls to the farmers to make sure they have received their monetary incentives and inquire if they are facing any challenges. Fig. 2 shows the \emph{AdSurv} communication model.
\\
The communication strategy was a follows
\begin{itemize} 
   \item Call the farmers every week to inform them of their number of reports and their reward to be posted on mobile money.
    \item Run a brief interview to find out what challenges they were facing; technical, network, social, schedule of their day to day activities in regards to the collecting of crop surveillance data an
    \item Log the information into a database and produce a weekly report to share with the stakeholders in \emph{NaCRRI and UNFFE}.
    \item For technical issues that were ease to resolve, the call centre members directed them to the software engineering team, who in turn contact the farmer directly for a follow up session.
    \item Offer over-the-phone tutorials to farmers on how to use the modules of the AdSurv app; collect images, automated diagnosis, access information on crop disease and pest, and ask the experts.
    \item Coordinate communication with district team leaders, who in turn support every cohort of farmers within their own district.
\end{itemize} 

A detailed section seven highlights the challenges and comments captured by the weekly routine calls.

\section{ Deployment and Training model}
%Travel routes and schedules
While two routes were pursued to optimise verification, for training and deployment, the field training team decided to run a unified one-team one-route sequential program. The justification was to have uniformity in the delivery of training materials to the trainees. Secondly the trainers needed to adapt and reform the training materials to the different communities and change the delivery of the content to match the hands-on training.
\\
The trainers in total covered seven training centres starting with the central at the \emph{NaCRRI} training centre, then near east was held in Mbale district, then far East in Soroti district and greater Northern sub region was covered in Lira district at the Ngetta ZARDI.
The West Nile was handled at the Abi ZARDI in Arua district, the near near West region held in Masindi district and finally the far Western region in Rwebitaba ZARDI held in Kabarole district.
\\
One day was allotted for each training centre, one day allocated for travel. The Table 1 shows some interesting statistics on the trainees per training centre.

\subsection{Training model}
%— Below in figure 5,  we show the training program that we followed for our daily training routine.
%i. Introductions, PEER, AdSurv, NaCRRI, Farmers 
%ii. Agronomy by NaCRRI - cassava, pest, disease, control
%iii. Introduction to mobile smart Phone - call, messages, Take an IMAGE
%iv. Break 
%v. Mobile Money 
%vi. Interact with the AdSurv app
%vii. Modules - Collect, Diagnosis, Q\&A, Chat, Channels
%viii. Lunch 
%ix. Field test 

\subsubsection{Training}
The training program was designed with the following objectives,
\begin{itemize}
    \item Introduce smallholder farmers to the cassava diseases and pest vectors.
    \item Familiarise the agents to operating a smartphone, taking an image, accessing their mobile money services.
    \item Train the farmers to use the \emph{AdSurv} app and conduct a practical field exercise.
\end{itemize}

Two training manuals were developed for the agronomy section and one for the mobile AdSurv sections. 
Figure 6 shows snippets of the manuals. Full manuals can be found on the following link: …………. 
[link/citation] %%%%%%%

\subsubsection{Peer-to-Peer learning methodology}
At every training centre were farmers who were already conversant with using a smartphone and the apps.
During the first section of the training, the trainers quickly identified farmers who were more conversant with operating mobile apps and the mobile smart phone.
Then the farmers were grouped into smaller training teams and distributed around the identified "tech-suvy" farmers, who then guided their colleagues during the training programme and practical field exercise. 
These "tech-suvy" quasi-trainers often interpreted the manuals into common local languages that were more understandable to the farmers from their regions.
\\
This farmer peer-to-peer training methodology achieved higher throughout compared to the centralised training methodology that is presented in the first pilot [citation] Mutembesa et al., 2018.

\subsection{ District farmers leader hierarchy.}
We grouped farmer trainees from each district and requested them to vote a district farmer leader who would be able to help the group member when challenged by using the application.
Each district had representative.
The district farmer leader methodology was adopted because of the infeasibility for expert trainers to making regular field visits to retrain and help farmers who weren't familiar using a mobile smartphone with common operational challenges. 
Some of these challenges include but not limited to; proper fill-in in of data collection forms, tuning on/off GPS, login into the application, asking questions using Q\&A, using the smartphone camera to take a good image etc, were easily handled by the district farmer leaders who many were "tech-sury" and conversant using the \emph{AdSurv} app.

\subsubsection{Responsibilities of district farmer leaders}
\begin{itemize}
   \item Assist in operating the \emph{AdSurv} app for assigned farmers agents in their region.
    \item Retraining exercise for farmers that missed on some modules of interest. 
    \item Ask questions on behalf of their fellow district farmers to agriculture experts using Chat and Q\&A module.
   \item Interpret and disseminate critical communication on application, updates and the reward structures.
\end{itemize}

\section{Incentive Structure}
For the scaled community sensing trial, we implemented a baseline monetary-based incentive for the participants. Weekly micro-payments were made out to the farmers who uploaded submissions through their mobile money accounts.\\
Mobile money is a service provided by telephony service companies, where mobile sim card (GSM terminal) mimics banking services. The user can receive, deposit, transfer and send money with their phone.
\\\\
The payments were calculated in batch as follows;
\begin{itemize}
    \item 1st payment scheme implemented in from mid April to end of May;\\
    Pay UGX 5000 (approx. USD 1.4) when agent completes uploading their first batch of 20 submissions. The rationale behind this is that an effort of 20 geo-tagged images is warrants the mobile money charges which are covered by the requester to sending the monetary incentive to the farmer.
    The requester also considers that UGX 5000 is a reasonable initial compensation for the effort, time and resources spent by the farmer to collect 20 images, valuing each report at UGX 250.\\
    After that, each of the next batch of 40 images will each be multiplied by the report value incremented cumulatively by UGX 25.
    
    \item 2nd payment scheme implemented in months June and July;\\
    Same as the 1st payment scheme only that after 400 reports, the batch size is increased to 50 reports for calculating the incremental cumulative reward.
    By this time the high performing members were reporting well over 400 reports per week and thus skewing both the compensation budget and data collection exercise. This was to encourage them make weekly submissions within 400.
    
    \item 3rd payment scheme implemented in month of August ;\\
    Same as 2nd payment scheme, but here the cumulative increment on the report price was capped at the 500th report. Reports after that were all calculated with a uniform UGX 500 per report.
    
    \item 4th payment scheme implemented from September to November;\\
     Same as 3rd payment scheme but all payments were capped at the 800th report and the batch size was increased to 100 for the cumulative increment in report price.
     
     \item 5th payment scheme implemented from November to December;\\
     This was the same as 4th payment scheme but due to budget exhaustion at the end of project, for all submissions beyond 500 images, the requester offered a flat UGX 200,000.
    
\end{itemize}

\section{Discussion of Results}
Over 227 days, the network of agents collected over 87,000 images from disparate locations around the country as show in Figure 5.\\
Most of the data was collected by farmers in the near East sub-region, followed by the central region, then North Eastern sub-region, followed by Northern region, then near Western region, then far West region and finally the North West Nile region as shown in Figure 4.
\\\\
The farmers submitted most of the images during the first four months of the project, with the highest submissions in the months of June, July and August 2018. % as shown in Figure 7.
These are the same times where drastic changes were made for the payment scheme iterating from 2nd scheme through to 5th payment scheme.
Submissions of more than 2500 images were coming in from a few individuals per week.
\\
The trend there after went on a downward trend because of budget exhaustion and anticipating the end of the project in November 2018 as seen in Figure 5.
\\\\
 Figure 8 shows the type of image data that was collected. The \emph{AdSurv} data collection module allows the agent to complete several micro-tasks and one of them is to label the image to which of these four categories it belongs ( Image labels; Disease, Whitefly, Anomaly and Other.)
\\
Most of the farmers reported images labelled as "Disease", followed by "Whitefly", then "Others" and lastly labels of "Anomaly" being the least collected.
\\ After the training, this makes the farmers aware of the problems facing their crops and this becomes a useful knowledge that helps them to identify diseased crop in the local farms they survey easily while at the same time sensitising fellow local farmers. 

\begin{figure}[htbp]
\centerline{\includegraphics[width=8.5cm]{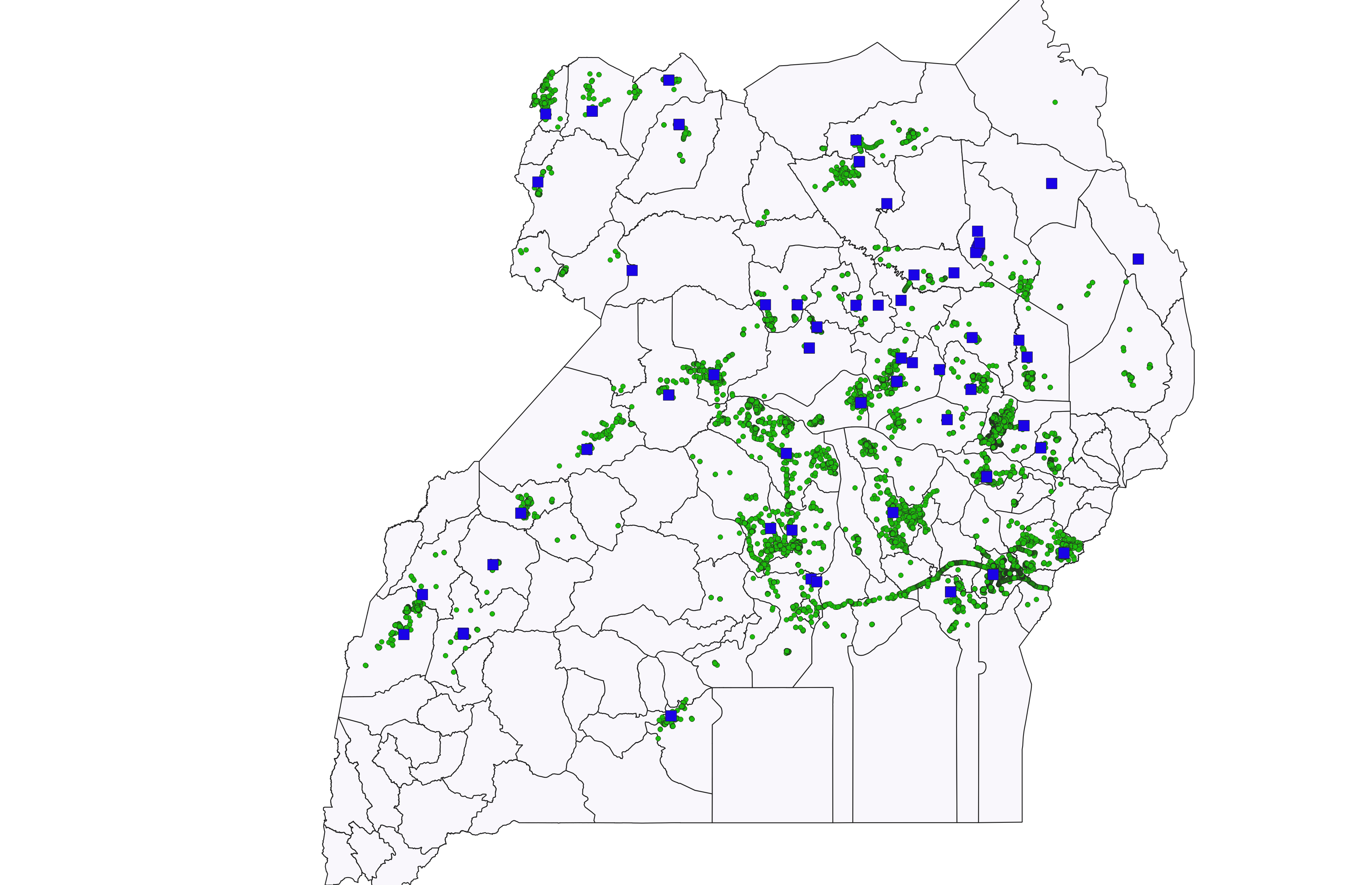}}
\caption{Map of Uganda showing farmer agents (blue) and data collected (green).}
\label{fig}
\end{figure}

\begin{figure}[htbp]
\centerline{\includegraphics[width=8.5cm]{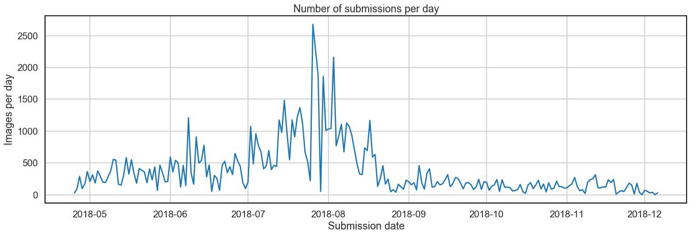}}
\caption{Reporting trends over the time span of the project}
\label{fig}
\end{figure}

\begin{figure}[htbp]
\centerline{\includegraphics[width=8.5cm]{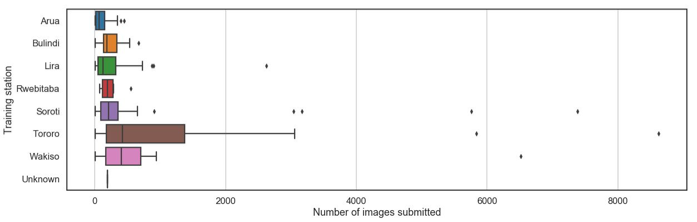}}
\caption{Reporting by regional training station \emph{ZARDI}}
\label{fig}
\end{figure}

%\begin{figure}[htbp]
%\centerline{\includegraphics[width=8.5cm]{/Users/macbook/Downloads/Publications/Conference-LaTeX-template_7-9-18/images/reporting-clusters.png}}
%\caption{Map of reports, that shows reporting clusters for famers around the regional training stations.}
%\label{fig}
%\end{figure}

\subsection{Characterisation of farmer agents}
%1. General Reporting behaviours \\\\
%2. Payment for the the eight months of pilot \\\\
%3. Characteristics exhibited by groups of agents (by gender, by age group, by district/region) \\\\
%4. Agent group statistics on usage of each module \\\\
%5. Explanation and Reason for usage patterns\\\\
\subsubsection{Spatial reporting trends}
Figure 9 shows the spatial density of reports submitted during the run of the pilot. From this we can infer knowledge on spatial reporting radius of a cluster of smallholder farmers within a given region.
\\ It is evident that the reporting trends tend to be further clustered within their own districts and around their surrounding districts.
This supports the proposition of clustered reporting behaviour of smallholder farmers that was highlighted in Mutembesa et al., 2018.
\\
From Table 1, we notice that while the farmers in the Northern region who were the highest in number, they were not the highest contributors. This was latter attributed to the farmers reporting crop loss in their gardens, due to prolonged dry season, which made it difficult to tell the difference between the diseased and healthy plants.

\subsubsection{Reporting by gender}Figure 8 shows the reporting by gender groups. The males submitted up to 67\% of the data while females contributed the rest. This is not surprising since this disparity is already exhibited int the recruited cohort of the volunteer smallholder farmers with a $ 64:36 $ recruitment ratio of $male : female$ respectively.
\\
However, the farmer with the highest number of submissions for the entire project was a female agent with close to 9000 image submissions. She was also a district team leader of her district and was a "tech-suvy" peer-to-peer trainer during the training session. Unlike other three female district team leaders, this female district team leader in the Eastern region influenced female farmers in her region to submit high number of reports.
This accounts for the near East region being the highest reporting sub region for the pilot lifetime.

\subsubsection{Reporting by age}
Figure 9 shows the reporting by the age groups, with the two age groups of 20-30 and 31-40, contributing the highest number of submissions, followed by the age groups of 41-50, then 51-60.
This goes to justify why one of the distribution model criteria during the sourcing of the farmer crowds included two youth per district to be identified in the selection process.
\\
One plausible reason for the first age group (20-30) demonstrating good reporting is they are mobile and often offer their labour on several farms, while the next age group (30-40) are mostly farmers that have families, own gardens and thus have greater social influence and capital within communities.
This means that they can easily be trusted to survey fields within their localities and their fellow farmers can approach then for advice.

%\subsection{Influence of monetary incentive }

\begin{figure}[htbp]
\centerline{\includegraphics[width=8.5cm]{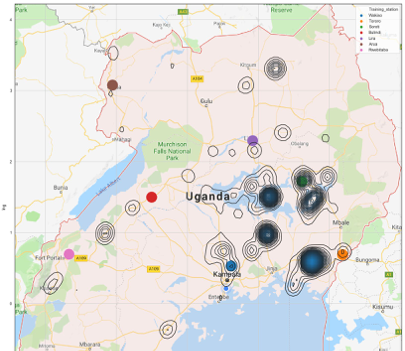}}
\caption{Spatial reporting densities. Where are most reports coming from?}
\label{fig}
\end{figure}

%\begin{figure}[htbp]
%\centerline{\includegraphics[width=8.5cm]{images/gender-reporting.png}}
%\caption{Reporting by gender}
%\label{fig}
%\end{figure}

\begin{figure}[htbp]
\centerline{\includegraphics[width=8.5cm]{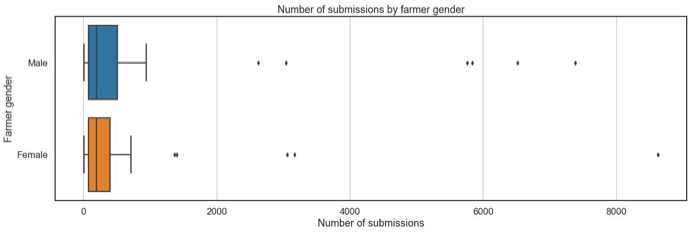}}
\caption{Submissions by gender}
\label{fig}
\end{figure}

\begin{figure}[htbp]
\centerline{\includegraphics[width=8.5cm]{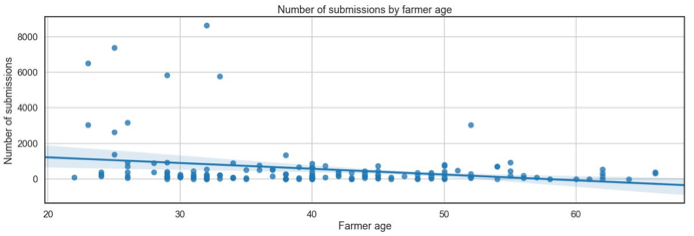}}
\caption{Reporting by age groups}
\label{fig}
\end{figure}

\subsection{Insights into farmer diagnosis}
While working with rural communities to gather information on crops, one question of interest to the national agricultural research organisation is the extent of disease and pest knowledge the smallholder farmers at training and through their participation in the campaigns. One way to assess this farmer knowledge is to correlate farmer observation completed as a micro-task in \emph{AdSurv} and diagnosis given by the subject matter expert.\\
The research attempted to categorise farmer comments according to frequently occurring diseases and symptoms and variations on their spelling and then created the following columns and encoded them with a 1 or 0.

The following categories including both a diagnosis and symptoms were identified and if any comments have and of the following keywords.
Keywords: ’leaves', 'root', 'stem','yellow', 'pale', 'cmd', 'cbsd', 'cbb', 'whitefly', 'cgm', 'chlorosis',
       'stunted', 'black\_spots', 'twisted', 'engulfed', 'folded', 'wilting',
       'curling', 'dry', 'rotten', 'lesions', 'candlestick', 'disease', 'pest',
       'anomaly', 'unhealthy', 'healthy', 'variety', 'other', 'unknown'
\\\\
%\subsubsectionThe Analysis
Of the 15,500 images annotated by the experts, considered images were those that had been reviewed by experts and also had a farmers diagnosis(comment) - currently around 7,491 images.
\\
The analysis considered the expert diagnosis as the ground truth, represented here as the “Actual”, and the 
Farmer comment as their diagnosis, represented here as the “Predicted”.
So far I have compared the primary expert diagnosis with the primary farmer diagnosis only. 
Below is a confusion matrix with predicted diagnosis along the columns and actual along the rows. Diagonals show number of accurate predictions

\begin{table}[htbp]
\caption{Farmer selection statistics} 
\begin{center}
\begin{tabular}{|c|c|c|c|c|c|}
\hline
\textbf  &\multicolumn{5}{|c|}{\textbf{Predicted}} \\
\cline{2-6} 
\textbf{Actual} & \textbf{\textit{CBB}}& \textbf{\textit{CBSD}}& \textbf{\textit{CGM}}& \textbf{\textit{CMD}}& \textbf{\textit{None}}\\
\hline
CBB & 655 & 736 & 15 & 110 & 2 \\
\hline
CBSD & 124 & 572 & 37 & 74 & 4 \\
\hline
CGM & 58 & 241 & 229 & 286 & 21 \\  
\hline
CMD & 217 & 488 & 435 & 3072 & 76 \\
\hline
None & 9 & 19 & 2 & 9 & 0 \\
\hline
all & 1063 & 2056 & 718 & 3551 & 103 \\
\hline
%Wakiso & Central & 7 & 36 & 23 & 13 \\
%\hline
%Totals &  & 48 & 175 & 112 & 63 \\
%\hline
%\multicolumn{4}{l}{$^{\mathrm{a}}$Sample of a Table footnote.}
\end{tabular}
\label{tab 2}
\end{center}
\end{table}

As expected CMD is most frequently and accurately defined. 
CGM very poorly identified by farmers.
CBB slightly better indicating that 62\% of plants identified as being CBB actually were CBB, while only 43\% of those actually having CBB were correctly identified.
CBSD is interesting because a high recall value indicates that 71\% of CBSD infected plants are correctly identified by farmers BUT with low precision, so only 28\% of those identified as CBSD by farmers actually turned out to be so.

\section{ Challenges and lessons}
%1. Call centre\\\\
%2. Limitations of monetary incentives with smallholder farmers.
\begin{itemize}
\item Inadequate training.
The time and resource allocation challenges significantly affected the quality of the regional training in-field with the farmers. The training team had very limited time to adequately cover all the modules of the \emph{AdSurv} app, the agronomy tutorials and the run 

\item Smartphone use. Since 75\% of the farmers were unfamiliar with smartphone use prior to the training, it was even harder for the team members on the call-centre to understand the problems faced by some of these individuals. One needed to do a very low-level analysis of the information given by the farmer in order to ascertain if they could indeed use the application or phone. More than 70\% of these instances revealed that the farmer didn’t actually know how to use the phone, load airtime or data bundles using mobile money, as well as use the \emph{Adsurv} application

\item Improper support. Despite the fact that the pilot came up with several strategies to help the farmers with some of their technology-related issues, as well as knowledge on cassava diseases were in place, the team at times faced constraints including funds and manpower.

\item When working with a large cohort of volunteers, it becomes intractable to follow the changes in strategies of participants. There is need for an automated and adaptive incentive mechanism that could make micro-payments in real-time in relation to the information being gained from the reports collects.
\end{itemize}

\section{Contribution}
This paper makes several contributions in the following ways;

\begin{itemize}
    \item The pilot presented here is the first of it kind structured community sensing campaign on scale, using mobile phones to collect crop surveillance with smallholder farmers in a low-resourced developing nation. The set and models presented here are replicable for purposes of surveillance of several phenomena from remote rural settings for example pest and vector surveillance, harvest monitoring, garden profiling and livestock tracking. 
    
    \item The paper also presents detailed models for crowd selection, communication and training of a volunteers at scale within their unique low-resource setting. It more importantly presents insights in how to design user-desired mobile apps for farmers.
    
    \item The paper also presents interesting insights into the behaviours of female volunteers under intrinsic incentives of leadership within their communities. This are important insights since the previous pilots suffered due to low participation of females on the projects.
    
    \item The communication model gives for such a set up allows a decentralised knowledge and information access for both the farmer and expert communities.
    
    \item The mobile application platform set was unique to allow both farmer and expert communication. In the mobile smartphone app were modules that allowed the expert to post new and simplified awareness materials on agronomy, disease and pest management, while allowing farmers directly ask questions to the experts. This bridged the gap between the two groups of users.
    
\end{itemize}

\section{Conclusion}
This paper demonstrates the viability of mobile phone based community sensing with local volunteers for monitoring phenomena that is important to those underserved rural remote communities.
It presents a set up vital for extending the gains presented here to other community sensing campaigns in various domains like agriculture, health, education etc.
\\
Future work building on this needs to tackle key optimisation challenges given the resource constraints and develop incentive mechanisms intuitive enough to adaptively factor the changes in seasonality of agriculture in relation to sensing objectives.
\\
Furthermore, the work presented here sets a unique ground to build platforms that are both user and expert facing, where the two sides can interact in a decentralised manner, about the subject matter of interest that they are sensing. The authors foresee this work contributing many important results to early warning systems for low-income developing world countries.

\section{Acknowledgements}
The authors would like to appreciate the generous funding from NSF/USAID grant and to the volunteer farmers without whom this work would never have been possible.

\section{References}

Mutembesa, D.,  Omongo, C., and Mwebaze, E. 2018.
"Crowdsourcing Real-Time Viral Disease and Pest Information: A Case of Nation-Wide Cassava Disease Surveillance in a Developing Country." In Proceedings of the Sixth AAAI Conference on Human Computation and Crowdsourcing, HCOMP 2018, Zurich, Switzerland, July 5-8, 2018. pages 117-125
\\\\
Agapie, E.; Teevan, J.; and Monroy-Herna ́ndez, A. 2015. Crowdsourcing in the field: A case study using local crowds for event reporting. In Third AAAI Conference on Human Computation and Crowdsourcing.
Chatzimilioudis, G.; Konstantinidis, A.; Laoudias, C.; and Zeinalipour-Yazti, D. 2012. Crowdsourcing with smart- phones. IEEE Internet Computing 16(5):36–44.
\\\\
Chklovski, T., and Gil, Y. 2005. Towards managing knowledge collection from volunteer contributors. In AAAI Spring Symposium: Knowledge Collection from Volunteer Contributors, 21–27.
\\\\
Chklovski, T. 2003. Learner: a system for acquiring commonsense knowledge by analogy. In Proceedings of the 2nd international conference on Knowledge capture, 4–12. ACM.
\\\\
Haklay, M. 2013. Citizen science and volunteered geo- graphic information: Overview and typology of participa- tion. In Crowdsourcing geographic knowledge. Springer. 105–122.
\\\\
Kanefsky, B.; Barlow, N. G.; and Gulick, V. C. 2001. Can distributed volunteers accomplish massive data analy- sis tasks. Lunar and Planetary Science 1.
\\\\\
Kittur, A.; Nickerson, J. V.; Bernstein, M.; Gerber, E.; Shaw, A.; Zimmerman, J.; Lease, M.; and Horton, J. 2013. The future of crowd work. In Proceedings of the 2013 conference on Computer supported cooperative work, 1301–1318. ACM.
\\\\
Mohan, P.; Padmanabhan, V. N.; and Ramjee, R. 2008. Nericell: rich monitoring of road and traffic conditions using mobile smartphones. In Proceedings of the 6th ACM conference on Embedded network sensor systems, 323–336. ACM.
\\\\
Nuwamanya, E.; Baguma, Y.; Atwijukire, E.; Acheng, S.; and Alicai, T. 2015. Competitive commercial agriculture in sub saharan africa. International Journal of Plant Physiol- ogy and Biochemistry 7(2):12–22.
\\\\
Otim-Nape, G.; Bua, A.; Thresh, J.; Baguma, Y.; Ogwal, S.; Ssemakula, G.; Acola, G.; Byabakama, B.; et al. 2000. The current pandemic of cassava mosaic virus disease in east africa and its control. The current pandemic of cassava mosaic virus disease in East Africa and its control.
\\\\
Paolacci, G.; Chandler, J.; and Ipeirotis, P. G. 2010. Running experiments on amazon mechanical turk.
\\\\
Quinn, A. J., and Bederson, B. B. 2011. Human computation: a survey and taxonomy of a growing field. In Proceedings of the SIGCHI conference on human factors in computing systems, 1403–1412. ACM.
\\\\
Rana, R. K.; Chou, C. T.; Kanhere, S. S.; Bulusu, N.; and Hu, W. 2010. Ear-phone: an end-to-end participatory urban noise mapping system. In Proceedings of the 9th ACM/IEEE International Conference on Information Processing in Sen- sor Networks, 105–116. ACM.
\\\\
Silvertown, J. 2009. A new dawn for citizen science. Trends in ecology \& evolution 24(9):467–471.
\\\\
Singh, P.; Lin, T.; Mueller, E. T.; Lim, G.; Perkins, T.; and Zhu, W. L. 2002. Open mind common sense: Knowledge acquisition from the general public. In OTM Confederated International Conferences” On the Move to Meaningful Internet Systems”, 1223–1237. Springer.
\\\\
Ssekibuule, R.; Quinn, J. A.; and Leyton-Brown, K. 2013. A mobile market for agricultural trade in Uganda. In Proceedings of the 4th Annual Symposium on Computing for Development, 9. ACM.
\\\\
Stevens, M., and DHondt, E. 2010. Crowdsourcing of pollution data using smartphones. In Workshop on Ubiquitous Crowdsourcing.
\\\\
Thiagarajan, A.; Ravindranath, L.; LaCurts, K.; Madden, S.; Balakrishnan, H.; Toledo, S.; and Eriksson, J. 2009. Vtrack: accurate, energy-aware road traffic delay estimation using mobile phones. In Proceedings of the 7th ACM conference on embedded networked sensor systems, 85–98. ACM.
\\\\
Van Pelt, C., and Sorokin, A. 2012. Designing a scal- able crowdsourcing platform. In Proceedings of the 2012 ACM SIGMOD International Conference on Management of Data, 765–766. ACM.
\\\\
Zhu, H.; Kraut, R.; and Kittur, A. 2012. Organizing without formal organization: group identification, goal setting and social modelling in directing online production. In Proceedings of the ACM 2012 conference on Computer Supported Cooperative Work, 935–944. ACM

\end{document}